# Brief overview on Bacteriophage therapy; Alternative to Antibiotics

Rameen Atique[1], Hafiza Arshi Saeed[2], Bushra Anwar[3], Tehreem Rana[4], Ayesha Haidar[5], Ayesha Muazzam[6], Areesha Naveed[7], Javeria Sharif[8], Aqsa Perveen[9], Hafiza Rida Fatima[10], Abdul Samad[11]*

**Affiliation Address**

1,2,3,5,7,8,9, 10 Department of Pathobiology and Biomedical Sciences, FVAS, MNS University of Agriculture, Multan Pakistan

[4]Department of Human Nutrition and Dietetics, MNS University of Agriculture, Multan Pakistan

[6]Department of Animal & Dairy Sciences FVAS, MNS University of Agriculture, Multan Pakistan

[11]Division of Applied Life Science (BK21 Four), Gyeongsang National University, Jinju 52852, Korea

**Corresponding Address**: buzdarabdulsamad@gmail.com

**Abstract:**

The term bacteriophage means killer or eater of bacteria. They were initially discovered by F.W. Twort and later on, Felix d'Herelle unveiled them to the world in 1910. Phage therapy has arisen as a favorable option to conventional antibiotics by reducing the multinational problem of increasing antibacterial resistance. These virulent viruses particularly prey on and contaminate bacterial strains and aid in fighting bacterial diseases. Researchers are performing various clinical trials on the bacteriophage to tackle pathogenic bacterial infections, varying from typical illnesses to highly invulnerable biofilms that cannot be treated with antibiotics. The primary experiments demonstrated that phage therapy has fewer consequences than traditional antimicrobial drugs. It is safer to use and show results within a few days. Although phage therapy has a wide range of promising results, but it also encounters diverse obstacles. One is that they are host-specific and can merely be used for personalized therapy. As thousands of bacteria can cause disease, clinicians have to construct a library of phage viruses. For successful treatment, an analysis of versatility, stability, and immune interference related to bacteriophage is necessary. Phage therapy is an excellent substitute for antibiotics as it illustrates a living base for the treatment of infections and it is climate-friendly. It only targets the pathogenic cells and has less influence on the normal microbiota. Regardless of the challenges and problems, phage therapy is approved as a beneficial approach to combating contagious infections.

**Introduction:**

Bacteriophages are dispersed in the environment making it astonishing that they were not found for nearly 40 years after the onset of research work in various laboratories in Europe. In 1896, Hankin documented that water in the rivers of India (Jumna and Ganga) had specific antibacterial action against many bacteria such as *cholera vibrio*. Ernest Hanbery Hankin attempted to count the number of bacteria (*vibrio cholera*) in cubic millimeters of water in the rivers of Ganga and Jumna. For this, he collected specimens where the river enters and leaves the Agra metropolis (Adhya & Merril., 2006; Kazhal & Iftimovich., 1968). Hankin enumerated around 1 lakh infectious organisms in water when it entered the city. Whereas the number of microbes is reduced to only 90 when water departs the city. This unusual process of self-cleansing of water in Ganghes was mentioned as "Hankin's phenomenon". In 1901, Emmerich and Löw noted that a molecule in an autolyzed living cell culture was adept at inducing lysis of different bacterial cultures to treat infections. Further studies on these strange organisms describe their action, impacts of bacterial toxins, and exhibition of lytic enzymes. All of these investigations were done on liquid cultures, but the first clear experiment on bacteriophage was operated on a solid culture medium. The chapter on modern phage research was initiated when Frederick W. Twort made an uncommon observation in 1915. During World War I, a scientist Felix d`Herelle uncovered a virus that could eat microbes. He prepared a phage medium culture to treat the warriors influenced by dysentery (Hausler, 2008). After that, it was reported that bacteriophages are non-toxic to humans and other creatures (Alessandrini & Doria., 1924; Compton, 1929; Sulakvelidze et al., 2001; Parfitt, 2005).

After the discovery of bacteriophages, d`Herelle utilized phages in clinical trials to deal with dysentery. The first attempt to use bacteriophage as a therapeutic agent was performed in Paris in 1919 (Summers, 1999). A boy with severe symptoms of dysentery was allocated d`Herelle anti-dysentery bacteriophages and he got well within a few days. In 1921, Richard Bruynoghe and Joseph Maisin employed bacteriophages to deal with staphylococcus skin infections. The phages are inoculated to the patients and the infections are suppressed within 24-48 hours (Rice, 1930; Schless, 1932; Stout, 1933). After the success of bacteriophage therapy for the treatment of dysentery, d`Herelle used distinct phage trials to cure many people in India affected by cholera and the bubonic epidemic (Kazhal & Iftimovich., 1968; Summers, 1999). Bacteriophages are the viral particles that are the natural terminators of bacterial cells. They take control of the cellular machinery of bacteria, duplicate themselves, and discharge new virus particles causing the bursting of bacterial cells (Clark & March., 2006). It is considered that there are approximately $10^{31}$ bacteriophages present on the planet Earth and they are important controllers of bacterial community in the biome (Suttle et al., 2005; Fuhrman, 1999). They play a vital role in maintaining normal homeostasis in living organisms and they exhibit characteristics to deal with bacterial disorders (De Paepe et al., 2014; Abedon et al., 2011).

The evolution of bacteriophage therapy is beneficial for mankind. Initially, antibiotics were useful to treat bacteria-associated infections, but many bacterial species have developed resistance against them. In this regard, lytic bacteriophages have proved possible substitutes for antibiotics (Padiyara et al., 2018; Cassir et al., 2014). The proteins present in the bacteriophages are widely used for the betterment of the community by managing bacterial infections, transduction, vaccine development, treatment for cancers, anti-biofilm agencies, and food conservation (Harada et al., 2018; Hsieh et al., 2017).

Many communities of micro-organisms are present in the ecosystem called microbiota. They help to establish relationships among different species and shape the biogeochemical pathways that occur in water and ground (Singer et al.,2017). Despite demonstrating relations with the water bodies and soil, these microbes can interact with multicellular living organisms such as plants and animals. They are essential in the maintenance of living bodies by inducing diseases and creating a symbiotic association with them. The microorganisms (bacteria, fungi, viruses) present in the gastrointestinal tract of humans cooperate in the process of digestion and enhancement of immunity (Belkaid & Hand.,2014; Sharon et al.,2016; Sonnenburg et al.,2005). Various parasitic associations take place in the gastrointestinal tract of humans where diverse microbes interplay with each other. The most significant parasitic relation in the gut is between bacteriophage (predators) and bacteria (prey). The viruses destroy the bacteria, and it results in the instabilities of both occupants (Faruque et al., 2005; Koskella & Brockhurst., 2014). A healthy human gut normally contains 35-2800 functional bacteriophage species and more than 50% of them are distinctive to each person (Manrique et al., 2016; Reyes et al., 2010). Although pages have been discovered for more than a hundred years their role in gut microbiota is less understood due to the lack of proper mechanisms for complementing the predator bacteriophages to their defined host. For this, the bacterial CRISPR technique is developed that harmonizes with the specific molecules of the viral gene pool (Paez-Espino et al., 2019).

Bacteriophages are perfectly suitable to be an ideal drug as it is favorably specific. Phages only prey on pathogenic microorganisms and spare the symbiotic microbiota. Because of their admiring specificity, bacteriophages are present ubiquitously in the environment and liquid solutions form a recess (Sandeep, 2006; Bruttin & Brüssow., 2005). They only lack the property of stability. In solutions, bacteriophages are partly stable because the virus particles modify their structure. This affects the storage of phage particles and decreases the shelf life of phage-built therapies (Gill & Hyman., 2010; Ackermann et al., 2004; Tovach et al., 2012; Jonczyk et al., 2011). Bacteriophages are widely used as therapeutic agents but their use in human clinical tests is restricted because of the following reasons: (1) phages are natural killers of bacteria but some of them may hold up bacterial vicious genes that can enhance the pathogenicity of bacterial species. Primary human pathogens that boost the virulence of bacteria are *Escherichia coli* and *Staphylococcus aureus* (Canchaya et al., 2003; Brüssow et al., 2004). Lytic phages are used instead of lysogenic phages to evade the threat of pathogenic bacteria. (2) Phages are utilized to transduce DNA from one infected bacterium to another. Because of the phage-mediated transduction, bacteria become tolerant to familiar bacteriophages like P1-like *Escherichia coli* and P-22-like *Salmonella.* To solve this, the number of bacteria and viruses should surpass the threshold level, and bacteriophage maladies should fit into the environmental conditions.

Since many species of bacteria have become resistant to antibiotics, phage therapy is an excellent replacement for antimicrobials. The seclusion of phage-associated lytic enzymes in phage treatment works likewise to the lysozyme enzyme in the antibiotics. Transmembrane protein holin and endolysin (cell wall hydrolase) play major roles in the breakdown of bacterial cells. The holin proteins deed as a clock in the lytic cycle and at the end of the cycle it turns on the entrance in the cell membrane that allows the endolysin proteins to access and hydrolyze the external membrane (Roach & Donovan., 2015). A newly discovered broad-spectrum phage lysin(endolysin) is ABgp46 which can lyse multi-drug tolerant bacteria comprising *Pseudomonas aeruginosa* and *Salmonella typhimurium* (Oliveira et al., 2016). Endolysins are more effective for bacterial decomposition than holins. They are harmless to multicellular organisms and can successfully eliminate bacterial pathogens. A combination of bacteriophage endolysins and antibacterial drugs can efficiently deal with bacterial infections (Wang et al., 2015). In addition, phage therapy provides more safety, specificity, and biofilm prevention in comparison to antibiotic treatment.

As we know d`Herelle discovered more than a hundred years ago that bacteriophage can be used to treat bacterial infections. Bacteria have become multiple drug-resistant and in this context, we need a promising treatment that can replace antibacterial resistance. Various trials have been done on bacteriophage therapy to make it a viable and universal exercise to investigate the framework of the antimicrobial resistance crisis and the possible solutions to tackle it (Podolsky, 2018). Different compounds like bacteriophages, antimicrobial drugs, and sulfonamides can inhibit the growth of bacteria. Penicillin, the first sulfa-drug, was honored as a drastic element in the removal of pathogenic diseases. The discovery of penicillin elevates the importance of antibiotics in the field of medicine. But in the 20th century, phage therapy raised its head in the domains of agriculture and pharmaceuticals. In the early stages, phage therapies were rejected to treat infectious diseases, but they were beneficial in genetic engineering products related to human and animal fitness and agri-food department in the early 2000s (Hausler, 2006; Kuchment, 2012). Bacteriophages are referred to as "industrial medicinal developments" that establish a global market through their regulatory demands of quality, protection, and efficiency. Phage therapy is applied broadly in Europe and standards are established that have the following outcomes: (1) to the grade product, all the processes and standards must be according to good manufacturing practices to produce bacteriophage. (2) phages should demonstrate their effectiveness and security in recombinant clinical trials. (3) phages must acquire marketing approvals.

The objective of phage therapy is to hinder infectious bacterial species without bothering the proportion of the natural microbiota of the patient. Bacteriophages continuously divide in the infected person to target the pathogenic bacteria with the help of lytic phage proteins. When phages arrive at the target site, they start to grow exponentially and kill bacterial cells to accomplish the known therapeutic consequence. After their function, phages are eradicated from the individual's body with urine (Zalewska-Piatek & Piatek., 2020; Strathdee et al., 2023; Nang et al., 2023). The use of phage therapy may cause ethical issues in some countries (Borysowski et al., 2019; Fauconnier, 2019; Zaczek et al., 2020). Page therapy includes an individual phage and a group of phages termed "phage cocktails" that are matched with the strain of bacteria in a patient to proceed with the treatment. Phage-based therapies and clinical experiments are performed in bacteriophage therapy headquarters that consists of the United States, Belgium, Georgia, and Australia (Knezevic et al., 2021; Lin et al., 2021; Suh et al., 2022).

Phage-based treatments are alternatives to antimicrobial therapies which are commonly used against bacterial diseases. It involves personalized therapy in which each patient is treated individually by the separation and estimation of specific infectious species of bacteria (Schooley et al., 2017; Terwilliger et al., 2020; Reuter & Kruger., 2020; Jones et al., 2022). Bacteriophage therapies are suitable for persons who have allergies to antimicrobial drugs, they are unable to expand in multicellular living cells and they provide safety measures to the patients (Ling et al., 2022; Loc-Carrillo & Abedon., 2011; Salmond & Fineran., 2015). In some cases, bacteria become immune to bacteriophages by mutations. Mutations cause changes in the genomic data of bacteria that develop resistance to phages (Lee et al., 2018; Fong et al., 2020). The development of phage drugs and therapies needs high-quality management standards.

**Discovery of Bacteriophages:**

(Duckworth, 1976) take part in explaining the discovery of bacteriophage viruses by discussing the historical background. F. W. Twort was an expert in the fields of pathology and bacteriology. He deals with the primitive causative entities responsible for causing diseases. He performed experiments to see the evolution of viruses in artificial culture media. As we all know viruses need living hosts to grow and multiply themselves, but Twort moved with his irrational assumption that viruses are the simplest form of life to exist on Earth. He believed that one time in the phylogenetic record, they must be able to live and inhabit an environment without living organisms. He made attempts to nurture viruses from soil, water, manure, and grass on specifically ready media. After that, he added some chemicals and extracts of fruits and vegetables to provide nutrition to the growing cells, but the results were negative. After the failure of this experiment, Twort inoculated an agar plate with some unfiltered liquid which is exclusively used for the immunization of smallpox. He noticed the growth of micrococcus bacteria in the agar medium and this bacterium (micrococcus) is associated with some diseases. It demonstrated watery areas in the inoculated agar. If these bacterial colonies are not grown in a different agar plate, then they evolve glassy or translucent. Twort documented the following interesting observations about the glassy modification of bacterial colonies: (i) The bacterial colonies impacted would not cultivate on any other culture medium. (ii) Investigation of transparent spots disclosed tiny granules and the absence of bacteria. (iii) If an isolated pure colony of micrococcus encountered the translucent part, then it made the entire colony glassy and clear. (iv) After the purification of the glassy component with the Chamberland candle persists in the ability to induce glazed adaptation. After all these observations, he deduced that the glassy modification was because of the contagious, filterable organism that can infect and kill bacteria in the cycle of multiplying itself. Later, this glassy compound was used to treat diseases caused by micrococcus.

**D`Herelle's disclosure of bacteriophages:**

Felix d'Herelle is the man who coined the name "bacteriophages" for bacteria-eating viruses. In 1910, when d'Herelle was in Mexico, when an attack of locusts happened around Yucatan. The ground was scattered with the remains of the infected locusts. He picked up a few of them who were suffering from black diarrhea. It is a disease caused by the bacteria coccobacilli. He prepared cultures of coccobacilli and after some time, he observed a strange behavior illustrated by the cultures. The bacterial colonies turned clear with glassy spots, round and 2-3mm (about 0.12) in width. After that, he observed the cultures under a microscope but there was nothing to see. After this huge discovery, d'Herelle worked on human dysentery. The isolated filtrates of viruses act on colonies of bacteria and agar plates are incubated. After incubation, he again noticed the clear patches. When he inoculated the filtered mixture of viruses in rabbits, they did not develop the disease. In the next experiment, he collected samples of stools from a patient suffering from bloody diarrhea and prepared an emulsion containing a few dips of stool purified from the Chamberland candle. d'Herelle inoculated the filtrate to a broth culture of dysentery bacteria (bacillus) and laid out a drop of the prepared blend on an agar plate. After the incubation period, the results were quite shocking for him, the test tube consisting of broth was clear because all bacteria had disappeared, and the agar plate had no bacterial growth. Bacteriophages are obligate parasites that grow in the living host for growth and nutrition.

**Mode of action of bacteriophage:**

(Dennehy & Abedon., 2021) discuss the mode of action bacteriophage and how the virus infects the bacterial cell and makes multiple virions. The life cycle of bacteriophage depends on several growth factors such as the time required for the adsorption of the virus particle, the length of lytic and lysogenic cycles, the number of new virus particles produced after the lysis of bacterial cells, and the stability of virions. Bacteriophage infection can be phage-productive in which there is the formation and release of newly produced virions, they may or may not use the bacteria and it indicates that the infected bacteria is responsible for chronic diseases. Phage-reductive infections are manifested by the survival of the phage gene pool and the shortage of new virus particles. In this type of infection, bacteria become lysogenized, and the bacteriophage is reduced to prophage (incorporation of bacterial chromosome with the viral DNA). Lastly, phage-destructive diseases are those in which there is no chance of survival for both the phage genetic material and phage offspring. When bacteriophage gains entry in the host bacterial cell, it induces infection by following two steps; which happens before the production of the primary vision and what will take place after that. It is the time during which either artificial lysis of phage occurs, or it remains intact with the bacteria. The recovery of bacteriophage from the artificial breakdown of bacterial cells is called the eclipse period. During this period, genes of the virus are expressed, and the genetic material of the phage is duplicated. Early genes (genes that express early during the eclipse phase) and late genes (genes that express in the later stage) are linked with the specific promoter regions in the phage genome. After replication, procapsids are the protein structures that surround the viral genome, and the viral components are assembled. Now, virion progeny is all set to release the host cell by keeping in check the beginning of lysis and the technique to destroy the outer protective wall of bacteria.

**Clinical trials of phage therapy:**

(Marongiu et al., 2022) highlight the fact that after the discovery of bacteriophage viruses, other scientists and researchers started to take an interest in the implementation of phage therapy. After World War II, bacteriophage therapy was used in the treatment of bacterial infections associated with animals. Over the years, phage therapy became popular in Poland where almost 550 persons were infected with bacterial diseases and their immune systems did not respond to antibacterial or chemical drugs. When they were administered bacteriophage therapy 546 patients showed positive results and only 4 out of 550 did not react to the therapy. Moreover, between the years 2008 and 2010, around 153 patients were admitted to the Phage Therapy Unit. Some of the cases demonstrated useful reactions to phage therapy but other patients obtained bacteriophage in combination with antibiotics. Patients became healthy in 6-65 days (about 2 months). The therapy contained 10-20 ml (about 0.68 oz) of the phage-containing solution at a congregation of $10^6$-$10^9$ phage forming unit/ml. The first clinical trial was conducted by utilizing the method of a randomized trial, placebo-regulated, and double-blind test. This study included 25 patients affected with *Pseudomonas aeruginosa* (burn injuries). Out of 25 cases, 12 were allocated to the remedy group and the remaining 13 to the control class. The former group was supplied with dressings drenched in a bacteriophage mixture (PP1131) and the latter control group was inoculated with an antibacterial drug, sulfadiazine. Patients included in the treatment group contained 12 distinct lytic phage viruses against *P. aeruginosa* with a titer value of $10^6$ PFU/ml. Also, cases involved in the treatment class were elder people and held up infectious pathogens that are more sensitive to antibiotic therapy. After 1 week, 6 out of 12 patients in the treatment level and 11 out of 13 cases in the control group were declared free from infection. Besides *P. aeruginosa, Staphylococcus aureus* and *Escherichia coli* were also considered in the clinical practices of phage therapy.

**Table 1: Outline of security and curative phage practices**

| Testing state | Investigational features | Aimed bacterium | Remarks | References |
|---|---|---|---|---|
| **Protective tests** | Animal examination | *E. coli* | Oral injection of T4 did not influence the microbiome of mice | Chibani-Chennoufi et al., 2004; Weiss et al., 2009 |
| | Human demonstration | Not applicable | Inoculation of T4 did not induce anti-T4 immunoglobulins | Bruttin & Brüssow., 2005 |
| **Stage (i)/(ii) clinical testing** | Phage treatment on endovenous leg sore | *E. coli, S. aureus, P. aeruginosa* | Any unfavorable results were not shown | Rhoads et al., 2009 |
| | Bacteriophage remedy in burn injuries | *P. aeruginosa, E. coli* | The presence of the host | Merabishvili et al., 2009; Rose et al., 2014 |

|  |  |  | bacterium stayed unaffected |  |
|---|---|---|---|---|
|  | The Polish medical records | Antimicrobial protective diseases | No severe outcomes; although the victory ratio was good | Weber-Dabrowska et al., 2000 |
| **Stage (iii) medical tests** | The Eliava examination: phage treatment in response to *Shigella* dysentery | *Shigella* species | An obvious decline in bacterial load was documented | Babalova et al., 1968; Sulakvelidze et al., 2001 |

**Safeness and effectiveness of phage treatment:**

(Speck & Smithyman., 2016) declared the safety and efficiency of phage therapy. In 1980, bacteriophages were administrated intravenously to deal with bacterial infections. In 1920, it was reported in the French and British review papers that intravenous phages could be a successful cure for typhoid. In 1931, Felix d'Herelle reviewed that IV phages could be used to treat cholera in India. In the 1930s, France and The United States successfully documented the use of intravenous bacteriophage antagonistic towards *Staphylococcus aureus*. Phage preparations are used in broth culture without peptone because peptone can induce shock. Researchers used bacteriophages to treat typhoid disease. In the first stage, the patients were injected with phages intravenously and they developed high fevers for some time. After that, the patients return to their original state rapidly. In the secondary phase, out of 56 cases, 53 were recovered from phage therapy, and the rest 3 were dead. The fatality rate was 5% (with bacteriophages) and 14% (without bacteriophages). Phage therapy is also useful in the treatment of Staphylococcal disorders. For this, bacteriophages are purified by CsCl a gradient separator at high speed, and a medium consisting of animal essence. In 1982, 48 patients acquired intravenous phages complementary to *S. aureus* suffering from lung or pleural infection. The ratio of the patients who received IV phages to the control group (without phages) are 95%:64%. A highly distinguished mixture of phages is injected intravenously to treat rhinosinusitis by removing *S. aureus* from the blood. Another safety measure elicited by phage therapy is the cure of pathogenic endocarditis. It is a disease of the internal areas of the heart including the chambers, caused by infectious *Staphylococcus aureus*. So, phage therapy is a possible remedy to decrease mortality caused by endocarditis.

**Phage therapy as an alternative to antibiotics:**

(Gordillo Altamirano & Barr., 2019) discussed the use of phage therapy in the modern world where microbes have become resistant to antibiotics. Treatment through antibiotics has been extensively used throughout history to save the lives of people. It was done by the expansion of medical therapies comprising organ implantation and cancer therapy. The development of antimicrobial resistance is a naturally occurring process where bacteria have modified themselves biologically according to modern-day antibacterials. Sometimes, bacteria automatically become immune to antibiotics due to the change in genes but sometimes, activities performed by humans develop antibacterial resistance in bacterial species. Bacteria have evolved mechanisms through which they have become multiple drug-resistant. It includes the horizontal transfer of genetic material from one bacterium to another through transformation (uptake of chromosomal material), conjunction (via pilli), and transduction (via bacteriophage). Bacteria can change the composition of the outer cell wall to inhibit the access of antibiotics in the cell. If the drug gains entry, bacterial cells alter it through the exhibition of enzymes like beta-lactamases. They can also terminate the productivity of enzymes that is essential for the triggering of antimicrobials. To tackle the problem of antibiotic resistance, bacteriophage therapy is used for treatment purposes by emerging rigid lytic phages. Phage therapy is responsible for the reduction in the activity of infectious pathogens and the disposal of contaminants and bacterial toxins. Phage therapy has earned importance in recent years because bacteria have not developed bacteriophage antagonism and the utilization of characteristic phage treatments that are unlikely to induce phages invulnerable to hosts. Bacteriophages are favorably precise for their hosts. Phages can "jump" and evolve with time and in the gastrointestinal tract, this strategy is stimulated by the microbiome. Phage therapy instantly targets infectious pathogens without disturbing the normal microbiota of the gut. Moreover, the early start of phage therapy is essential for its success because if there is a pause of almost 6 hours then there will be a reduction in the efficacy of therapy. The seclusion and compilation of well-distinguished bacteriophages from biological origins are typically sensual for phage medicine. Bacteriophages interconnect with the immune system of the host and are termed natural assassins because they decrease the mass of the bacteria in the environment. The function of immunity in this regard is to inhibit the population of bacteria for bacteriophage therapy to flourish. This association between phages and immunity is called "immunophage synergy". In clinical trials, phage therapy is used in combination with antibiotics called "phage antibacterial synergy". Antibacterial drugs and phage therapy have synergistic effects in reducing the bacterial population. It has benefits in restricting the use of antimicrobials to reduce bacterial resistance and deal with multidrug-resistant microbes. At the termination of the lytic cycle, bacteriophages activate different sets of enzymes to promote the bursting of bacterial cells and the discharge of virus particles. Holins are capable of permeability and drilling holes in the plasma membrane. The timing of action of holins depend on the specificity of genes or alleles. Later on, endolysins (lysins) using these holes undergo locomotion from the cytoplasm to the outer layer (cell wall). The process of cell lysis is facilitated by the action of protein spanins, causing catalysis of the bonding of internal and external walls. All of the discussed modifications make phage therapy a better replacement for antibiotics.

**Challenges and opportunities of phage therapy:**

(Pires et al., 2020) explains the challenges and future scope of phage therapy. The recent ultimatum associated with bacteriophage therapy is: (1) The manufacturing of phages should be under strict rules and regulations to guarantee its high-grade criteria. During production, the factors of lysogenic chromosomes, infectious proteins, antimicrobial opposition, and endolysins should be avoided. (2) A bacteriophage can be a prospective nominee for treatment if it has a good shelf life. The stability of phage mixtures relies on the threshold of bacteriophage in the territory of infection. If phage preparations are stored for longer periods then their efficiency in treating bacterial infections decreases. (3) Phage viewing techniques are devised to detect and quantify the bacteriophages through qPCR, flow cytometry, and surface plasmon resonance. (4) Biofilms are the aggregation of bacterial cells that are defended by a matrix. The matrix takes up phages provoking the proficiency of bacteriophages to agitate biofilms. Some enzymes act on the polysaccharides of the matrix and make way for phages to enter the bacterial cell. The formation of a nutrient gradient in the biofilm leads to the formation of the resting cells in the inner layers of the biofilm. When bacteriophages infect these dormant cells they are incapable of replicating and multiplying. Furthermore, recent studies have documented that the bacteria have developed resistance against phages. It has occurred because of the appearance of bacteriophage-insensitive mutants (BIMs). Bacteria acquire mutant genes that formulate resistance to phages. The developed resistance can be treated with a combination of different phages termed phage cocktails. Cocktails can decrease the evolution of BIMs because they are made of fixed arrangements comprising a wide range of hosts.

**Advantages of phage therapy:**

(Loc-Carillo & Abedon., 2011) talk about the merits of phage therapy. Bacteriophages are powerful bactericidal agencies. Once lytic phages infect the bacteria, they are incompetent to retain their feasibility. As phages are made up of genetic material (DNA and RNA) and proteins, they are not toxic. Phages are highly host-specific and that's why they can only infect infectious pathogens. They did not influence the normal microflora or bacteria. Since phage therapy is a recent treatment for infectious bacteria, they have a narrow ability to induce resistance. There are low chances of cross-resistance with antibiotics because both of them have different modes of action. They can damage bacteria in a single dose, are natural living substances, and can be deactivated by fluctuations in sunlight, heat, pH, and temperature. Lastly, bacterial populations are not susceptible to antibiotics but they can be destroyed by phages as they are biofilm clearer.

**Conclusion:**

Bacteria are ubiquitous in the ecosystem, some of them are beneficial for our health like normal gut microflora and some of them are virulent like biofilm-forming bacteria. Bacteriophages are viruses that infect the bacterial hosts and inhibit their growth. Due to the incline of antibacterial resistance in the bacterial species, phage therapy has gained much importance in the field of medicine and agriculture. Various clinical trials have been performed on humans and animals to enlighten the efficacy of bacteriophages *in vitro* but it has hurdles *in vivo* investigational practices. To lessen the chances of resistance and mutations in bacterial hosts, phage cocktail therapies are used. Phage therapy is considered biological medicinal outcomes that have a low shelf life and can modify themselves to target pathogenic bacteria. The inflexible prerequisite of bacteriophages as a medicinal derivative causes problems like ethical permissions related to therapy, inhibition of phage treatment in multidrug-resistant bacteria, and decline in the significance of investigation and production of new drugs. Since phage therapy has proved useful for ecological health, rules and regulations authorize different nations to adopt it as a replacement for antibiotic treatment.